\def\eeq{\end{equation}}
\def\beq{\begin{equation}}
\def\bea{\begin{eqnarray}}
\def\eea{\end{eqnarray}}

\documentstyle[prd,aps,floats]{revtex}

\begin{document}

\renewcommand{\topfraction}{1.0}
\twocolumn[\hsize\textwidth\columnwidth\hsize\csname 
@twocolumnfalse\endcsname

\title{Precision cosmology as a test for statistics}

\author{Diego F. Torres${^{1,2,}}$\thanks{dtorres@venus.fisica.unlp.edu.ar} }

\address{${^1}$Departamento de F\'{\i}sica, 
Universidad Nacional de La Plata,
C.C. 67, 1900, La Plata,  Argentina\\
${^2}$Astronomy Centre, CPES, University of Sussex, Falmer,
Brighton BN1 9QJ, United Kingdom}

\maketitle

\begin{abstract}

We compute the shift in the epoch of matter-radiation equality due to the
possible existence of a different statistical
(non-extensive) background. The shift is mainly caused
by a different neutrino-photon temperature ratio. We then consider the prospects
to use future 
large galaxy surveys and cosmic microwave background measurements
to constrain the degree of non-extensivity of the
universe.


\end{abstract}

\vskip2pc]

\subsection{Introduction}

The possible  
need of non-extensive thermostatistics in cosmology, gravitation and
astrophysics has been discussed by several authors, for instance see Refs. 
\cite{1c}.
Indeed, the question of to what extent 
the universe as a whole is an extensive system is well
justified. It is known that whenever long range 
interactions are present, standard
(Boltzmann-Gibbs)
statistical 
sums and integrals may diverge (see Ref. \cite{TSALLIS-REVIEW} for a review). 
In any case, to check the validity of a theory we must locate it in a more 
general framework,
a group of test theories, 
which in general will have a free parameter. Then, one has to isolate
which of the theories of the group are able to reproduce 
experimental or observational facts.
Although this is a common practice in cosmological models based on
alternative theories of gravity, it is not so usual concerning alternative
statistics. The problem in these cases is 
try to determine if different statistical theories
may also yield the same observational results.

The group of test theories will be, 
for our purposes, that introduced by Tsallis in \cite{first}. By use of them we
expect to check the validity of the Boltzmann-Gibbs scenario.
Let us recall the starting points of nonextensive statistic (NES), 
in Tsallis' approach. The formalism begins by postulating \cite{first}:

\noindent {\bf Postulate 1}.- {\it The entropy of a system that can be found
with probability
$p_i$ in any of  $W$ different microstates $i$ is given by}

\beq
S_q=k\frac {1}{(q-1)}\, \sum_{i=1}^W [p_i\,-\,p_i^q]=
k \frac {1}{q-1} \left( 1-\sum_{i=1}^{W} p_i ^q\right), 
\label{Sq}
\eeq
{\it with  $q$ a real parameter}. We have a different statistics for
every
possible $q-$value. In (\ref{Sq}) we have used, of course, that

\beq
\sum_i p_i =1.
\eeq
In general,

\beq
\label{Sq2}
S_q=k \frac{1-{\rm Tr} \rho^q}{q-1}.
\eeq

\noindent {\bf Postulate 2}.- {\it An experimental measurement of an observable
$A$,
whose
expectation value in microstate $i$ is $a_i$, yields the $q$-
expectation
value} (generalized expectation value (GEV))

\beq
\label{gev}
<\,A\,>_q\,=\,\sum_{i=1}^W\,p_i^q \,a_i = {\rm Tr} \rho^q \hat A,
\eeq
{\it for the observable} $A$.

These two statements
have the rank of {\it axioms}. As such, their 
validity is to be decided exclusively by the
conclusions to which they lead, and ultimately by comparison with
observations.
We shall not present any further
the formalism in this work, referring the reader to  
Refs. \cite{TSALLIS-REVIEW,first,Curado,TIRNAKLI}. 
A more detailed presentation can also be found in \cite{TORRES_V}.
Instead we shall use only the results we need for our 
calculations. 
This non-extensive formal approach was applied in many different systems 
including Newtonian gravity \cite{Plas1,Plas2}, Levy type diffusion
\cite{Ale-Z}, the problem of solar neutrinos \cite{solar} and others.
Concerning the universe as a whole, we have previously shown that 
this group of test 
theories will affect primordial nucleosynthesis yields and the evolution of the
neutron to baryon ratio 
in a computable way.
Later, this 
could  be used to set bounds upon the free parameter 
they have and thus the upon
the degree of non-extensivity present in the early universe 
\cite{TORRES_V,TORRES_PRL,TORRES_U}. In addition,
a generalization of the Planck radiation
law was derived and applied to the cosmic microwave background data 
\cite{TSALLIS_RAD,P2V}.
In all these works,  limits upon the non-extensive (free) parameter 
were imposed. The typical order of magnitude of the bounds is
$10^{-4}$, what means that possible Bolztmann-Gibbs' 
deviation may be about one part in ten thousands.

\vspace{0.5cm}

In this letter we shall consider 
the matter-radiation equality epoch. This moment of cosmic history 
 is important for every cosmological model because
this transition changes the growth rate of density perturbations. Whereas
perturbations inside the horizon are frozen during a radiation dominated universe,
perturbations on all scales can grow (and structure can form) during the matter
era \cite{Lucchin}. Then, the size of the horizon at the time of
the matter-radiation equality is an important parameter of the model (indeed, it is
the only one for CDM models) and it 
is to be read from the spectrum of density
perturbations \cite{Lucchin}. In the following years, new satellites will 
measure cosmic
microwave anisotropies and new large galaxy surveys will be avalaible. In
particular, the Sloan Digital Sky Survey (SDSS) 
will acquire 10$^6$ redshifts \cite{TEGMARK,Stra}.
This will make cosmology to enter in an era of high precision and about ten
cosmological parameters will be accurately determined \cite{Jung}.
In what follows,
we expect to show that this set of data can also be used as a test of the
underlying statistical theory. To do so, 
we shall compute  the shift in the matter-radiation equality
that would be 
caused by non-extensivity; in general,  by the 
use  of a  different statistical
background.  
As this changes the shape of the power spectrum of density perturbations, 
something that can be
measured from galaxy surveys
\cite{TEGMARK,GOL}, a new bound upon the statistical description will arise.
In addition, we shall comment on the prospect for the accurate determination of
the cosmic neutrino background and the effective number of neutrino species.
This will also allows us to extract conclusions on which statistical
model may be the right description of the universe.

\subsection{Non-extensive corrections and matter-radiation equality}

The photon energy density 
is dominated by the cosmic
microwave background. Its temperature is measured extremely well \cite{FIX},

\beq
T_{\gamma,0}=(2.728 \pm 0.004 ){\rm K}
\eeq
and this gives, for the standard model,

\beq
\rho_{\gamma,0}=4.66 \times 10^{-34} \;\;{\rm g\;\;cm^{-3}}.
\eeq
However, if we are to take into account possible non-extensive effects, the
energy density will in general be given by,

\beq
\rho_q=\frac{\pi^2}{30} gT^4 + \frac1{2\pi} \left( 40.02 g_b + 34.70 g_f \right) T^4
(q-1),
\label{den}
\eeq
where $g_{b,f}$ are degrees of freedom of bosons and fermions,
$g=g_b+7/8g_f$ and $q$ is a real free parameter measuring the amount of
non-extensivity in the system \cite{TORRES_V,TORRES_PRL}. When $q=1$,
all results reduce themselves to the standard ones.

The present radiation density also has a contribution from relic neutrinos that,
due to the impossibility of detecting them directly, must be fixed theoretically.
The relationship between neutrino and photon temperature was computed in
Ref. \cite{TORRES_V} for a non-extensive framework and is given by,

\beq
T_\nu=T_\gamma \left(\frac 4{11}\right)^{1/3} \left(1+ (q-1) 0.013 \right).
\label{8}
\eeq
As we shall see below, equation (\ref{8}) is an important relationship, with a number of
observational consequences.
If $N_\nu$ is the number of neutrinos families (hereafter considered as three), then

\beq
\rho_\nu=\frac 78 \frac{\pi^2}{15} T_\nu ^4  N_\nu 
\left[ 1 + \frac{15}{\pi^4 } \frac 87 34.70 (q-1) \right],
\eeq
which, adding to $\rho_\gamma$ obtained from (\ref{den}) leads to,

\beq
\rho_{rad}= \frac{\pi^2}{15} T_\gamma ^4   
\left[ 1 + 0.681 + 10.33 (q-1) \right]
\eeq
for the total radiation density.
The first term in the bracket is the usual contribution of photons, the second
stands for the usual contribution of neutrinos while the third sums up all
non-extensive corrections.

The Friedmann equation for a zero curvature
Friedmann--Robertson--Walker
universe is

\beq
\left(\frac{\dot a}{a}\right)^2=\frac{8\pi G}{3} \rho,
\eeq
where $a$ is the cosmological scale factor and $\rho$ is the energy density,
which has contributions of non-relativistic
matter and radiation. If these fluids are
non-interacting, the conservation equation $T^{\mu\nu}_{\;\;\;;\nu}=0$ holds for
them separately and one gets,

\beq
\rho_m=\rho_{m,0} \left(\frac{a_0}{a}\right)^3, \hspace{2cm}
\rho_{rad}=\rho_{r,0} \left(\frac{a_0}{a}\right)^4.
\eeq
Here, $\rho_{m,0}$ stands for the present matter density (which is fixed if we
assume spatial flatness) and $\rho_{rad,0}$ for the analogous radiation density.
The redshift in the matter-radiation equality is given by,

\beq
1+z_{eq}= \frac{a_0}{a_{eq}}=\frac{\rho_{m,0}}{\rho_{rad,0}},
\eeq
Then, 

\beq
1+z_{eq}= \frac{ 3H_0^2}{8\pi G} \frac{1}{\rho_{rad,0}}=24000h^2 \left[
1-\frac{10.33}{1.68} (q-1) \right],
\eeq
where $H_0$ is the present value of the Hubble constant and $h$ is the Hubble
constant in units of 100 km s$^{-1}$ Mpc$^{-1}$. Without using the full solution
for the matter-radiation era, existing here because of the validity of General
Relativity, it is possible to estimate the Hubble radius at equality by assuming
that the matter solution holds all the way backwards in time.  At equality,

\beq
\frac{a_{eq} H_{eq}}{a_0 H_0}=\sqrt{2} \sqrt{1+z_{eq}}=219h\left[1+(1-q) 3.08
\right],
\eeq
where it was used a matter dominated solution for the scale factor in a flat
Friedmann-Robertson-Walker universe, $a(t)\propto t^{2/3}$, equivalently, the
dependence of $\rho_m$ given in (12). The $\sqrt{2}$ factor appears because at
equality, the right hand side of equation (11) has two equal energy densities.
Again, the first
term in the bracket is a complete standard result while the second one is the
correction due to the change in the statistical model. 

\subsection{Galaxy surveys and parameter estimation}

Now the analysis proceeds in much the same way as was done in Ref. \cite{A}
for Brans-Dicke gravitation, which can be seen for details. The shift in the
matter-radiation equality will lead to a shift in the maximum of the power
spectrum of horizontal nature. Such a shift was considered by Tegmark
\cite{TEGMARK}
who introduced a phenomenological parameter $\eta$ and considered how
accurately it can be measured by SDSS.
Tegmark gives two expected estimates on $\Delta \eta /\eta$,
(see panel 3 and 4 of Fig. 1 of Ref. \cite{TEGMARK}).
 The first one corresponds to the case in which all parameters
are considered fixed --or measured by other experiments-- except $\eta$. The
second one, to the case in which all parameters are to be extracted from SDSS
alone. For the scale currently going non-linear ($k \sim 0.1h {\rm Mpc}^{-1}$),
$\Delta \eta /\eta  \sim 0.02$ and  $\Delta \eta /\eta  \sim 0.1$ respectively
\cite{TEGMARK}. The accuracy of the measurements on $\Delta \eta /\eta$
depends on the scale considered. The scale taken here --and also in \cite{A}--
avoid one to worry about non-linearity or bias effects. 
These estimations can even be
improved considering data of other satellites, like MAP or Planck, as explained in
\cite{TEGMARK}.

In the normal General Relativity situation, the focus is pointed towards the
determination of $\Omega_mh$, where $\Omega_m$ is the matter density
parameter. Changing $\Omega_m h$ yields a horizontal shift to the
power spectrum.
In our situation, using $\Delta \eta /\eta \sim 0.02$ and equation (15)
we can read limits upon $q$ to be of order
10$^{-3}$ which is comparable to the nucleosynthesis bound obtained in
\cite{TORRES_PRL,TORRES_U} and slightly less restrictive than
 the more in depth analysis
of \cite{TORRES_V} and the bounds given in \cite{TSALLIS_RAD} and 
\cite{P2V}.
However, it comes now from a different arena, being the properties of galaxy
surveys and the neutrino-photon temperature relationship its main inputs.

\subsection{Detectability of the cosmic neutrino background}

In the standard Boltzmann-Gibbs statistics, the ratio of the energy density of 
neutrinos to that of photons is, as we have noted before,

\beq
\left(\frac{\rho_{\nu}}{\rho_{\gamma}}\right)_{BG}
= \frac 78 \left(\frac 4{11} \right)^{4/3}
N_\nu =0.681.
\eeq
However, recent studies have pointed out that the assumption that neutrinos
decoupled completely before $e^+-e^-$ annihilation is not entirely correct
\cite{DICUS,GNEDIN}. If the neutrinos share in the heating somewhat, their
energy density is larger than the above quoted value. This can be modeled
as if the number of neutrinos families $N_\nu$ is effectively 
bigger than 3. Together
with QED corrections to the energy density of $e^{\pm}$ and $\gamma 's$
\cite{HECKLER}, the increase in the effective number $N_\nu$ is,

\beq
\delta N_\nu=0.04 \;\;\;{\rm to}\;\;\;0.05.
\eeq
This leads to an increase of $\sim 1\%$ 
in the neutrino energy density.

In the non-extensive context, we may think of formula (9) as an effective variation
in $N_\nu$, given by

\beq
\delta N_\nu=\frac {15}{\pi^4} \frac 87 34.70 (q-1)=6.10 (q-1).
\label{nvar}
\eeq
Then, if a direct detection of the cosmic neutrino background would be avalaible,
a straightforward 
comparison with (\ref{nvar}) may also yield a bound upon $(q-1)$.

The precision detection of the neutrino energy density was considered by Lopez
et al. \cite{LOPEZ}. They analized the possible consequences that a small
increase in $\rho_\nu$ would have in the cosmic microwave background.
New measurements planned for the forthcoming satellites MAP and Planck lead
them to show that the sensitivity will be so great that the neutrino energy density
will be detectable by itself.
As in the work by Tegmark, estimates come in two different ways. If all other
parameters (for instance, baryon density $\Omega_B$, Hubble constant $H_0$,
slope of the primordial perturbations $n$, etc) are held fixed (as given by other
experiments), the expected sensitivity in $N_\nu$ is given by Fig. 3 of Ref.
\cite{LOPEZ}.
Going up to multipole moment $l \sim 1000$ and using only temperature
anisotropy data, the expected sensitivity will be $\sim 0.01$ \cite{LOPEZ}.
If the standard result $N_\nu=3$ is confirmed, one can then obtain --using
(\ref{nvar})--
a bound upon $(q-1)$ of order  $\times 10^{-3}$. This could even be improved
using
polarization data in the analysis, as explained in Ref. \cite{LOPEZ}.

\subsection{Concluding remarks}

We have presented the case for obtaining 
new cosmological bounds upon the degree of non-extensivity
of the universe. These new bounds will come from accurate comological
measurements of the microwave background and galaxy surveys.
To allow for this bounds to be obtained,
we have computed the shift in the epoch of matter-radiation
equality due to a different statistical background.
Afterwards, 
comparing with the expected measurements of $\Delta \eta
/ \eta$, we have shown how this can be used to constrain
the possible non-extensive parameter. 
We have also analized the implications that a precision detection of the cosmic
neutrino background may have concerning the fixing of the statistical description
and we have shown that another bound may arise from there.
The order of magnitude of these two bounds is expected to be $|q-1| 
\leq 10^{-3}$ and can be
compared with other previously obtained cosmological constraints
\cite{TORRES_PRL,TSALLIS_RAD}.
However it is important to stress here
that many phenomena can cause the same effect. For instance, in the galaxy
survey oriented scheme,
 a change in $\Omega_m$ or $h$,
a different number of masless species or a change in the theory of gravity to a 
scalar-tensor one have all the same output on $a_{eq} H_{eq}$. 
In the case of the correction to the effective number of neutrino species $N_\nu$,
it may well be the case that the 1\% correction arising from the extra heating
and QED effects be hidden by corrections arising from a value of $q$ different
from 1. To determine this, the former corrections need to be computed directly in
the non-extensive framework, without assuming Boltzmann-Gibbs statistics as
an initial hypothesis. 
It is disturbing
to note that whereas there is only one way of being {\it standard}
there are many different and arbitrary ways to deviate from that situation.
Even more disturbing is to see that different deviations may produce
the same observational output.
Despite of these caveats,
observational facts
like the one discussed here, as well as those concerning nucleosynthesis, 
are in the way of determining if the usual statistics is the right description of the
universe. Although it of course seems a rather safe assumption, and in particular
values of $|q-1|>10^{-3}$ are completely discarded, there is certainly the need
for more work to get a definite answer.

\subsection*{Acknowledgments}

During the course of this research, the author was a Chevening Scholar and 
acknowledge partial support from CONICET (ARGENTINA)
and The British Council (UK).
He wishes to acknowledge A. Mazumdar, A. R. Liddle and U. Tirnakli
for useful comments.

\end{document}